\documentclass[groupedaddress, superscriptaddress, prl, showpacs,  amsmath, amssymb, preprint]{revtex4}  
\usepackage{graphicx, color}

\begin{document}

\title{Laboratory observation of a nonlinear interaction between shear
  Alfv\'{e}n waves}
\author{T.A. Carter}
\email{tcarter@physics.ucla.edu}
\author{B. Brugman}
\affiliation{Department of Physics and Astronomy, University of California, Los Angeles, CA 90095-1547}
\affiliation{Center for Multiscale Plasma Dynamics, University of California, Los Angeles, CA 90095-1547}
\author{P. Pribyl}
\author{W. Lybarger}
\affiliation{Department of Physics and Astronomy, University of California, Los Angeles, CA 90095-1547}

\begin{abstract}
An experimental investigation of nonlinear interactions between shear
Alfv\'{e}n waves in a laboratory plasma is presented.  Two Alfv\'{e}n
waves, generated by a resonant cavity, are observed to beat together,
driving a low frequency nonlinear psuedo-mode at the beat frequency.
The psuedo-mode then scatters the Alfv\'{e}n waves, generating a
series of sidebands.  The observed interaction is very strong, with
the normalized amplitude of the driven psuedo-mode comparable
to the normalized magnetic field amplitude ($\delta B/B$) of the
interacting Alfv\'{e}n waves.

\end{abstract}


\maketitle

The Alfv\'{e}n wave is the fundamental low-frequency normal mode of a
magnetized plasma and is a ubiquitous feature of space plasmas
({\itshape e.g.} the auroral ionosphere~\cite{GH84} and the solar
wind~\cite{BD69}) and laboratory plasmas ({\itshape e.g.}
tokamaks~\cite{HS93} and linear devices~\cite{G94}).  The linear
properties of these waves have recently been explored in detailed
laboratory experiments~\cite{LG99, KB03, VGM04}.  However, the
nonlinear behavior of Alfv\'{e}n waves has not been
investigated in the laboratory. From a weak turbulence point of view,
nonlinear interactions between Alfv\'{e}n waves are responsible for
the cascade of energy in magnetohydrodynamic (MHD)
turbulence~\cite{K65}.  In the incompressible MHD approximation an
anisotropic energy cascade results from interactions between
counter-propagating shear Alfv\'{e}n waves~\cite{SM83,GS97}. In MHD
turbulence theories, interactions are generally assumed to be local in
wavenumber space and density fluctuations are assumed to play only a
passive role in the cascade~\cite{BN01}.  However, nonlocal
interactions between shear waves can generate beat-wave driven density
fluctuations.  This mechanism is essential in decay instabilities such
as the parametric~\cite{HC76,SL96} and modulational~\cite{WG86,H94}
decay instabilities, where the pump and daughter Alfv\'{e}n waves beat
together to drive an ion acoustic wave or a nonlinear psuedo-mode,
respectively.  In addition, density fluctuations are an integral part
of Alfv\'{e}n waves with small perpendicular scale (dispersive kinetic
or inertial Alfv\'{e}n waves~\cite{VGM04}) and can therefore become
active participants in the cascade as it approaches the dissipation
scale~\cite{terry01}.  Density fluctuations generated at small scale
can scatter large scale Alfv\'{e}n waves~\cite{DM01}, and could
therefore influence the cascade at larger scales.

In this Letter, an observation of a nonlinear interaction between
shear Alfv\'{e}n waves in a laboratory plasma is presented. Large
amplitude shear Alfv\'{e}n waves are generated using a resonant
cavity.  In circumstances where two waves are simultaneously emitted
by the cavity, production of sideband waves and low frequency
fluctuations at the sideband separation frequency is observed.  The
interaction is identified as a beat-wave interaction between
co-propagating shear waves, where a nonlinear psuedo-mode is driven at
the beat frequency.  The psuedo-mode then scatters the Alfv\'{e}n
waves, generating a spectrum of sidebands. The amplitude of the driven
pseudo-mode is substantial, much larger than would be predicted by
simple magnetohydrodynamic (MHD) theory.

The experiments were performed in the upgraded Large Plasma Device
(LAPD), which is part of the Basic Plasma Science Facility
(BaPSF)~\cite{lapd} at UCLA. LAPD is an 18~m long, 1~m diameter
cylindrical vacuum chamber, surrounded by 90 magnetic field
coils. Pulsed plasmas ($\sim 10$~ms in duration) are created at a
repetition rate of 1~Hz using a barium oxide coated nickel cathode
source. Typical plasma parameters are $n_e \sim 1 \times 10^{12}$
cm$^{-3}$, $T_e \sim 6$~eV, $T_i \sim 1$~eV, and $B < 2$kG.  The
experiments were performed using helium as a working gas.  LAPD
plasmas have values of $\beta = 2\mu_o n k_{\rm B} (T_{\rm e} + T_{\rm
  i})/B^2$ comparable to the electron-to-ion mass ratio, $\beta \sim
m_e/M_i$ (typical $\beta$ values range from $5\times 10^{-5}$ to
$1\times 10^{-3})$.  In these experiments, $\beta \gtrsim m_e/M_i$,
and the electron thermal speed is therefore larger than the Alfv\'{e}n
speed ($v_{\rm th,e} > v_{\rm A}$).  The ion sound gyroradius,
$\rho_s$, in these experiments is $\sim 0.5-1.5$~cm.

\begin{figure}[!htbp]
\includegraphics[width=3.0truein]{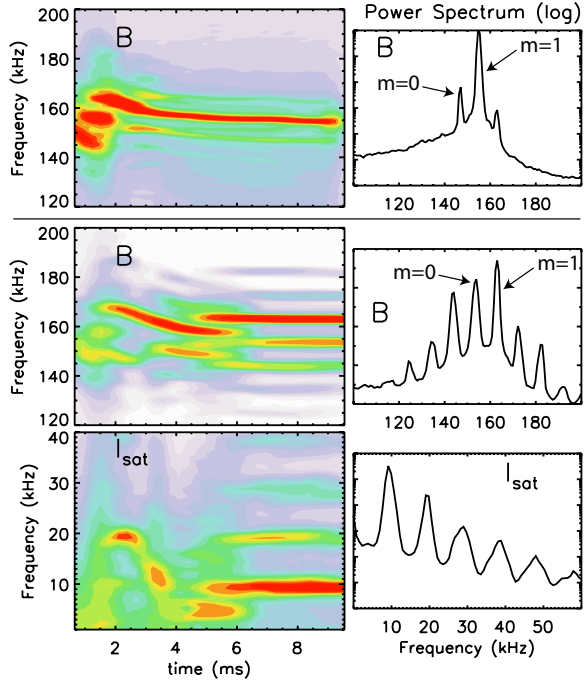}
\caption{Top panel: typical power spectrum of shear Alfv\'{e}n waves
during MASER emission.  Bottom:  $B$ and $I_{\rm sat}$
power spectra during MASER emission with the floating plate closed.}
\label{figure1}
\end{figure}

Large amplitude shear Alfv\'{e}n waves are generated using the
Alfv\'{e}n wave MASER~\cite{mm03,mm05}. The nickel cathode and
semi-transparent molybdenum anode of the plasma source in LAPD define
a resonant cavity from which spontaneous shear Alfv\'{e}n wave MASER
emission at $f \sim 0.6 f_{\rm ci}$ is observed, with $k_\perp \rho_s$
typically $0.3-0.5$.  The emitted shear waves are eigenmodes of the
cylindrical LAPD plasma column, with $m=0$ or $m=1$
azimuthal mode number.  The amplitude of the waves can be as
large as $\delta B/B \gtrsim 1$\% and can be controlled through
changing the plasma source discharge current.  Emission of the waves
typically begins early in the LAPD discharge, as the source region
current is ramping up and plasma parameters and profiles are evolving.
During this early phase the MASER emission is observed to ``mode
hop,'' where a sudden transition in the frequency of emission
is observed~\cite{mm05}.  The top panel of Figure~\ref{figure1} shows
the FFT (fast Fourier transform) power spectrum versus time for
magnetic field fluctuations measured in the main LAPD plasma column
($\sim 7$~m from the source region) along with a line plot of the
spectrum at $t=8$ms (helium discharge, 700G, $f_{\rm ci} = 266$~kHz).
Two mode hops are apparent during the early evolution of the magnetic
field spectrum, and the second hop is identified through mode
structure measurements as a transition from $m=0$ to $m=1$.  After the
mode hop, the $m=0$ mode does not completely disappear, but persists
at a much lower level as is apparent in the accompanying line plot.
An additional frequency is observed in the spectrum, located above the
$m=1$ frequency by the difference frequency between the $m=0$ and
$m=1$ modes.  This mode appears to be a sideband of the primary $m=1$
mode which could be generated by a weak nonlinear interaction between
the $m=0$ and $m=1$ mode.

It was found that much stronger sideband generation is observed during
high-current discharges when the length of the main LAPD plasma column
is shortened to 10~m by terminating the plasma with an
electrically-floating aluminum plate.  The bottom panel of
Figure~\ref{figure1} shows FFT power spectra of measured magnetic
field ($B$) and Langmuir probe measured ion saturation current
($I_{\rm sat} \propto n_{\rm e} \sqrt{T_{\rm e}}$) during MASER
emission with the floating plate closed.  The magnetic spectrum is
considerably more complicated in this case, showing fluctuations at a
number of discrete frequencies and an additional mode hop (which is
likely due to a change in radial mode structure).  In the steady state
phase of the discharge ($t > 6$~ms), there are two strong modes
present (identified as $m=0$ and $m=1$), surrounded by a number of
sidebands.  Simultaneously, low frequency $I_{\rm sat}$ fluctuations
are observed at the sideband separation frequency and harmonics. These
observations are consistent with the following scenario: (1)
Simultaneous emission of two large amplitude shear modes ($m=1$ and
$m=0$) from the cavity occurs. (2) The two shear waves beat together,
driving a low frequency fluctuation at the beat frequency. (3) The low
frequency fluctuation then scatters the incident shear waves, leading
to the generation of the observed sidebands.  Our conjecture is that
termination of the the high discharge current plasma with the floating
plate induces simultaneous spontaneous emission of large amplitude
$m=0$ and $m=1$ modes, leading to the proposed scenario.

\begin{figure}[!htbp]
\includegraphics[width=3.0truein]{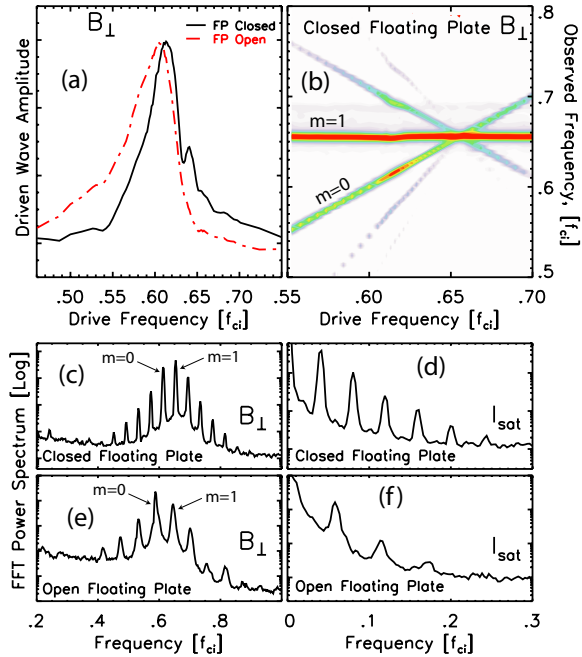}
\caption{(a) Cavity response as a function of normalized external
  drive frequency.  (b) Magnetic power spectrum due to interacting
  shear waves during a frequency separation scan. For drive frequency
  on the $m=0$ resonance: (c) Magnetic power spectrum and (d) ion
  saturation current spectrum with the floating plate closed.  (e)
  Magnetic power spectrum and (f) $I_{\rm sat}$ power
  spectrum with the floating plate open}.
\label{figure2}
\end{figure}

In order to conclusively determine the nature of the sideband
generation, the capability to externally drive the resonant cavity was
developed. To excite the cavity, oscillating currents are driven
between the anode and cathode using external power
supplies. Figure~\ref{figure2}~(a) shows the cavity response
(magnitude of the emitted shear wave, measured in the main plasma
column) as a function of drive frequency (normalized to $f_{\rm ci}$)
for cases with the floating plate open and closed.  The resonant
nature of the cavity emission is clear, as is the modification in the
properties of the cavity upon closing the floating plate.  The
external drive couples primarily to the $m=0$ mode, but the additional
peak observed with a closed floating plate corresponds to the $m=1$
mode frequency, and may indicate coupling to that mode.  The cavity
was then operated so that the spontaneous emission of only one
mode was observed ($m=1$), and a second mode ($m=0$) was externally
excited, allowing an investigation of the interaction between the two
shear waves with a variable separation frequency.  These experiments
were conducted with the floating plate both open and closed. Lower
discharge currents were necessary to limit the spontaneous cavity
emission to a single mode with the plate closed.

Figure~\ref{figure2}~(b) shows the magnetic power spectrum versus
drive frequency during a frequency separation scan with the floating
plate closed.  The spontaneous emission ($m=1$) is fixed during the
scan (the horizontal feature at $f \sim 0.66 f_{\rm ci}$) while the
frequency of the driven mode is changed.  The production of sidebands,
in particular the first upper sideband, is evident over a wide range
of frequency separations between the driven ($m=0$) and spontaneous
($m=1$) waves.  The change in sideband amplitude versus frequency separation
is primarily due to changes in the externally driven mode amplitude
with frequency (due to the $Q$ of the cavity).
Figure~\ref{figure2}~(c) shows a line plot of the magnetic power
spectrum when the drive frequency is at the peak of the $m=0$
resonance, $f_{\rm drive} = 0.61 f_{\rm ci}$.  A number of sidebands
around the driven ($m=0$) and spontaneous ($m=1$) shear waves are
visible.  Figure~\ref{figure2}~(d) shows the $I_{\rm sat}$ power
spectrum at the same drive frequency, which exhibits fluctuations at
the sideband separation frequency and harmonics.
Figure~\ref{figure2}~(e) and (f) show $B$ and $I_{\rm sat}$ spectra
during driven-spontaneous interaction with the
floating plate open. Shifts in the absolute frequency and frequency
separation of the $m=0$ and $m=1$ modes are evident relative to the
closed floating plate case.  Aside from these differences in the
linear cavity response, the observed spectra are quite similar to
those observed in the closed floating plate case: magnetic sidebands
are evident, as are low frequency $I_{\rm sat}$ fluctuations at the
sideband separation frequency.  This suggests that the primary role of
the closed floating plate in the data shown in Figure~\ref{figure1} is
to modify the cavity properties so that two large amplitude shear wave
modes are emitted simultaneously.  The mechanism by which the floating
plate causes this multi-mode emission is not yet understood, but it
is known that plasma parameters and profiles are altered relative to
the non-terminated case.  The presence of the floating plate also
allows for the possibility of reflection of the Alfv\'{e}n waves and
therefore the development of a counter-propagating population of
waves.  In addition, the closed floating plate and the anode define a
second resonant cavity which can support shear Alfv\'{e}n wave modes.
The qualitative similarity between the closed and open floating plate
cases shown in Figure~\ref{figure2} suggests that the second cavity
and the associated reflected shear waves are not essential for the
interaction, and that instead it is more likely an interaction between
co-propagating shear waves emitted by the cavity.

\begin{figure}[!htbp]
\includegraphics[width=3.0truein]{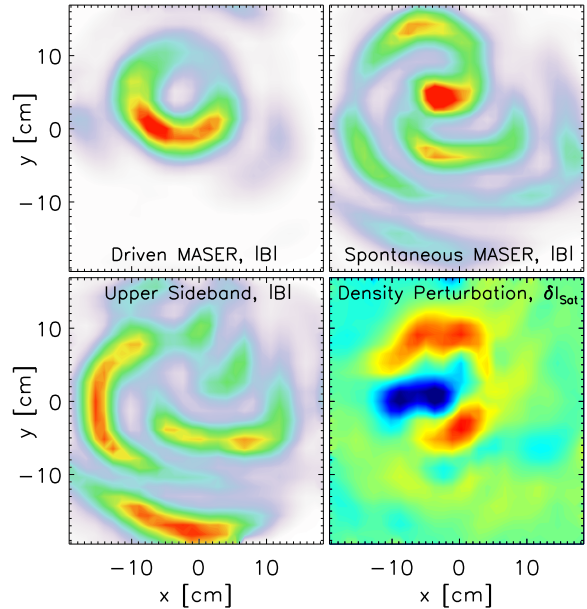}
\caption{Structure of the driven MASER, spontaneous MASER, upper
  sideband, and low frequency $I_{\rm sat}$ fluctuation in
  the plane perpendicular to $B$.}\label{figure3}
\end{figure}

Figure~\ref{figure3} shows the structure of the fluctuations during an
interaction between driven and spontaneously emitted cavity shear
modes.  Due to the shot-to-shot phase variation in the spontaneous
emission, cross-correlation techniques were used to determine the
structure of the interacting modes.  Two probes were used, one fixed
at one spatial location (to provide a phase reference) while the
second was moved shot-to-shot to 961 positions in the plane
perpendicular to the background field by computer control.  The two
probes were separated axially (along the magnetic field) by
approximately 1~m.  The driven MASER wave has an $m=0$ structure, with
a hollow amplitude profile and a single current channel.  The
spontaneous MASER exhibits primarily an $m=1$ structure, with two
counter-rotating current channels and a peaked amplitude profile
between the two.  The structure of the upper sideband has some
similarity to the spontaneous $m=1$ emission, but is more concentrated
in the periphery, which may be consistent with $m>1$ mode content.
The pattern of the $I_{\rm sat}$ fluctuation is more centrally
localized, and is not readily identifiable as any single azimuthal
mode number.

\begin{figure}[!htbp]
\includegraphics[width=2.5truein]{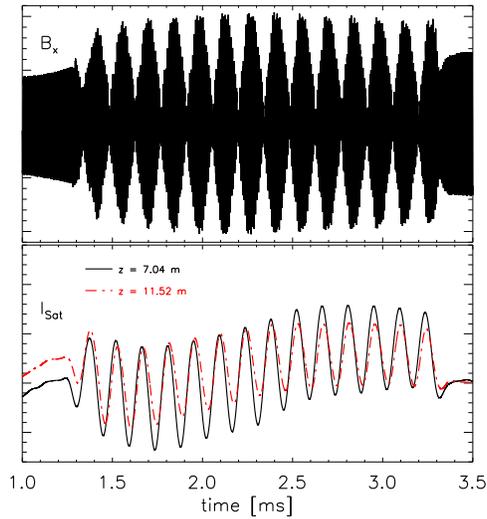}
\caption{Top: raw magnetic pick-up loop signal during beat wave
  experiment. Bottom: simultaneous $I_{\rm sat}$ measurements at two
  different axial locations.}\label{figure4}
\end{figure}

Figure~\ref{figure4} shows raw $B$ and $I_{\rm sat}$ fluctuation
measurements during an interaction experiment where both shear waves
are driven with the floating plate open ({\itshape i.e.}  there is no
spontaneous emission from the cavity).  In this case both modes are
$m=0$ and each driven slightly off resonance to achieve the desired
frequency separation.  The raw data shows that the beat wave
interaction also occurs in this case, indicating that the interaction
is not dependent on the presence of two different azimuthal modes.
Simultaneous measurements of $I_{\rm sat}$ at two axial locations
separated by $2.88$~m are shown from which a parallel phase velocity
for the low frequency fluctuation can be computed.  The average
parallel phase velocity in the region of constant plasma density is
measured to be $294 \pm 35$~km/s, where the Alfv\'{e}n speed is $\sim
550$~km/s and the ion acoustic speed is $\sim 13$~km/s.  The computed
phase velocity is consistent with three-wave matching rules ($\omega_1
+ \omega_2 = \omega_3$ and $\vec{k}_1 + \vec{k}_2 = \vec{k}_3$) using
the kinetic shear Alfv\'{e}n wave dispersion relation, $k^{2}_{\|}=
\omega^{2} \left(1-\omega^{2}/\omega^{2}_{ci} +
k^{2}_{\perp}\rho^{2}_{s} \right)$, where $k_{\perp}\rho_{s}$ is taken
from a spatial fit of the shear wave eigenmode structure to be $\sim
0.38$ for the two interacting Alfv\'{e}n waves~\cite{mm05}.  The low
frequency fluctuation does not correspond to a linear plasma wave
({\itshape e.g.} an ion acoustic wave) and is instead a beat-wave
driven nonlinear perturbation or psuedo-mode.  While the observations
are not consistent with a decay instability, it should be pointed out
that the modulational instability involves a similar set of modes: in
this instability, the pump shear wave decays into a
forward-propagating sideband wave and a pseudo-mode at the sideband
separation frequency.

The amplitude of the $I_{\rm sat}$ fluctuation is observed to scale
approximately bilinearly with the amplitude of the two interacting
shear waves.  The normalized amplitude is quite large with $\delta
I_{\rm sat}/I_{\rm sat} \sim 1-10\% \gtrsim \delta B/B$, substantially
larger than the normalized density perturbation which would be
predicted by simple ideal MHD theory ($\delta n/n \sim (\delta
B/B)^2$).  In addition to density, $I_{\rm sat}$ is sensitive to
electron temperature and to any population of fast electrons (those
with energies greater than the negative probe bias $\sim 65V$), and
therefore fluctuations in these quantities might explain the magnitude
of the observed $I_{\rm sat}$ signal. However, preliminary microwave
interferometer measurements indicate that there are significant
line-average density fluctuations associated with the psuedo-mode, and
that it is therefore largely a fluctuation in density.  Future work
will focus on measurements to more accurately determine the magnitude
of $\delta n/n$ in the psuedo-mode and compare it with more
comprehensive theoretical predictions.

In summary, a nonlinear beat-wave interaction between shear Alfv\'{e}n
waves has been observed.  Two resonant-cavity-produced, co-propagating
shear waves are observed to beat together, resulting in a low
frequency fluctuation at the beat frequency and the subsequent
creation of Alfv\'{e}nic sidebands.  The low frequency fluctuation is
identified as a nonlinearly driven psuedo-mode generated by the beat
between the two co-propagating shear waves. Counter-propagating
interactions will be explored in future experiments, where beat-wave
driven ion acoustic waves may be possible.

The authors would like to thank S.~C. Cowley, W. Gekelman, J.~E. Maggs,
and G.~J. Morales for invaluable discussions.  This work was completed
using the Basic Plasma Science Facility at UCLA, which is funded by
DOE and NSF.  This work was supported by DOE grant DE-FG02-02ER54688
and by DOE Fusion Science Center Cooperative Agreement
DE-FC02-04ER54785.

%\bibliographystyle{prsty}
%\bibliography{ref}

\end{document}